\documentclass[conference]{IEEEtran}
\usepackage{cite}
\usepackage{graphicx}
\usepackage[fleqn]{amsmath}
\usepackage{comment}
\usepackage{bm}
\usepackage{amsfonts}
\usepackage{multirow}
\usepackage{booktabs}
\usepackage{subfigure}
\begin{document}
  \bstctlcite{IEEEexample:BSTcontrol}

  \title{Transfer Learning-Based Received Power Prediction \\
  with Ray-tracing Simulation and Small Amount of Measurement Data}

  \author{
    \IEEEauthorblockN{
      \normalsize Masahiro Iwasaki\IEEEauthorrefmark{1},
      \normalsize Takayuki Nishio\IEEEauthorrefmark{1},
      \normalsize Masahiro Morikura\IEEEauthorrefmark{1}, and
      \normalsize Koji Yamamoto\IEEEauthorrefmark{1}\\
    }
    \IEEEauthorblockA{
      \IEEEauthorrefmark{1}\small Graduate School of Informatics, Kyoto University,
      Yoshida-honmachi, Sakyo-ku, Kyoto, 606-8501 Japan\\
      E-mail: nishio@i.kyoto-u.ac.jp
    }
  }
  \maketitle

\begin{abstract}
This paper proposes a method to predict received power in urban area deterministically, which can learn a prediction model from small amount of measurement data by a simulation-aided transfer learning and data augmentation. Recent development in machine learning such as artificial neural network (ANN) enables us to predict radio propagation and path loss accurately.
However, training a high-performance ANN model requires a significant number of data, which are difficult to obtain in real environments.
The main motivation for this work was to facilitate accurate prediction using small amount of measurement data. To this end, we propose a transfer learning-based prediction method with data augmentation.
The proposed method pre-trains a prediction model using data generated from ray-tracing simulations, increases the number of data using simulation-assisted data augmentation, and then fine-tunes a model using the augmented data to fit the target environment.
Experiments using Wi-Fi devices were conducted, and the results demonstrate that the proposed method predicts received power with 50\% (or less) of the RMS error of conventional methods.
\end{abstract}
  \IEEEpeerreviewmaketitle

\section{Introduction}
\label{sec:introduction}
Radio propagation has been investigated for a long period, and many path loss prediction models have been studied in literature. Traditionally, they are classified roughly into two categories: empirical and deterministic models. The empirical models such as Okumura-Hata \cite{okumura_hata}, Longley-Rice \cite{Longley_Rice}, and ITU-R P.1812 \cite{ITU_R}, describe the relationship between the power and the environment parameters from a statistical point of view. The empirical models are simple to implement and computationally efficient. However, there could exist environments where the models are not accurate without manual tuning.

On the other hand, the deterministic models can predict site-specific radio propagation. The deterministic models are derived from laws of physics, such as ray-tracing \cite{ray_tracing}, Maxwell equations \cite{Maxwell}, and radiosity \cite{radiosity}. Therefore, the models are complex to implement and computationally expensive, and their prediction accuracy depends on the accuracy and resolution of available maps depicting the environment such as hills, buildings, and trees.

Machine learning and especially artificial neural networks (ANNs) have been proposed to obtain prediction models that are more accurate than the standard empirical models while being more computationally efficient than the deterministic models. Many literatures show that the ANNs predict path loss and received power successfully \cite{imai, 7930454, 133098, 4022648, 1416569, 848409}. However, these schemes require large numbers of training data samples, which are obtained through measurement which requires significant human resources, money, and time.

Geographical interpolation based inference methods have been proposed to predict radio propagation accurately with smaller amount of data. Linear interpolation is the simplest way to infer unknown value from data measured in the vicinity. Prediction methods based on Kriging, which is developed originally in geostatistics, have been shown to perform received power and interference prediction successfully \cite{7127661, 6817823, 7817747}. Kriging assumes the interpolated values are modeled by a Gaussian process governed by prior covariances, but there exists an area that does not follows Gaussian process. For example, path loss drastically increases when the mobile station (MS) enters a building shadow. Kriging is difficult to capture such drastic change, while the deterministic models such as ray-tracing and ANN based methods with building map data \cite{imai} can capture.

This paper proposes an ANN based prediction method that infers received power from small amount of measurement data. The key idea is sim-to-real transfer, which pre-trains a prediction model with simulation data and fine-tunes the model with small amount of measurement data \cite{Sim_To_Real}. The proposed method learns site-specific characteristic of radio propagation roughly from ray-tracing simulation data, and fit the model to the real environment with the measurement data. Additionally, a data augmentation method based on simulation results is presented to increase the number of available data samples, thereby increasing the prediction accuracy of the fine-tuned model. The main contributions of this study are summarized as follows:
\begin{itemize}
\item 
We designed a framework to train a received power prediction model using transfer learning and data augmentation to facilitate accurate prediction with small amount of measurement data. Leveraging ray-tracing simulations based on 3D map data enables us to learn radio propagation models considering geographical characteristics and augment the measurement data by generating data samples artificially from original samples in consideration with site-specific characteristic shown in the simulated data.
\item 
We constructed an ANN model for received power prediction. Features representing long-term propagation characteristics and a feature representing short-term propagation characteristics are separately inputted into the constructed model to capture both characteristics fairly.
\item 
We experimentally demonstrate that the proposed scheme can predict received power more accurately than previous schemes. Outdoor experiments were conducted using off-the-shelf IEEE 802.11 devices in a 5\,GHz channel.
\end{itemize}

\section{Transfer Learning-Based Received Power Prediction}
\label{sec:proposed_scheme}
\subsection{System Model}
\begin{figure}[!t]
  \centering
  \includegraphics[width=0.5\textwidth]{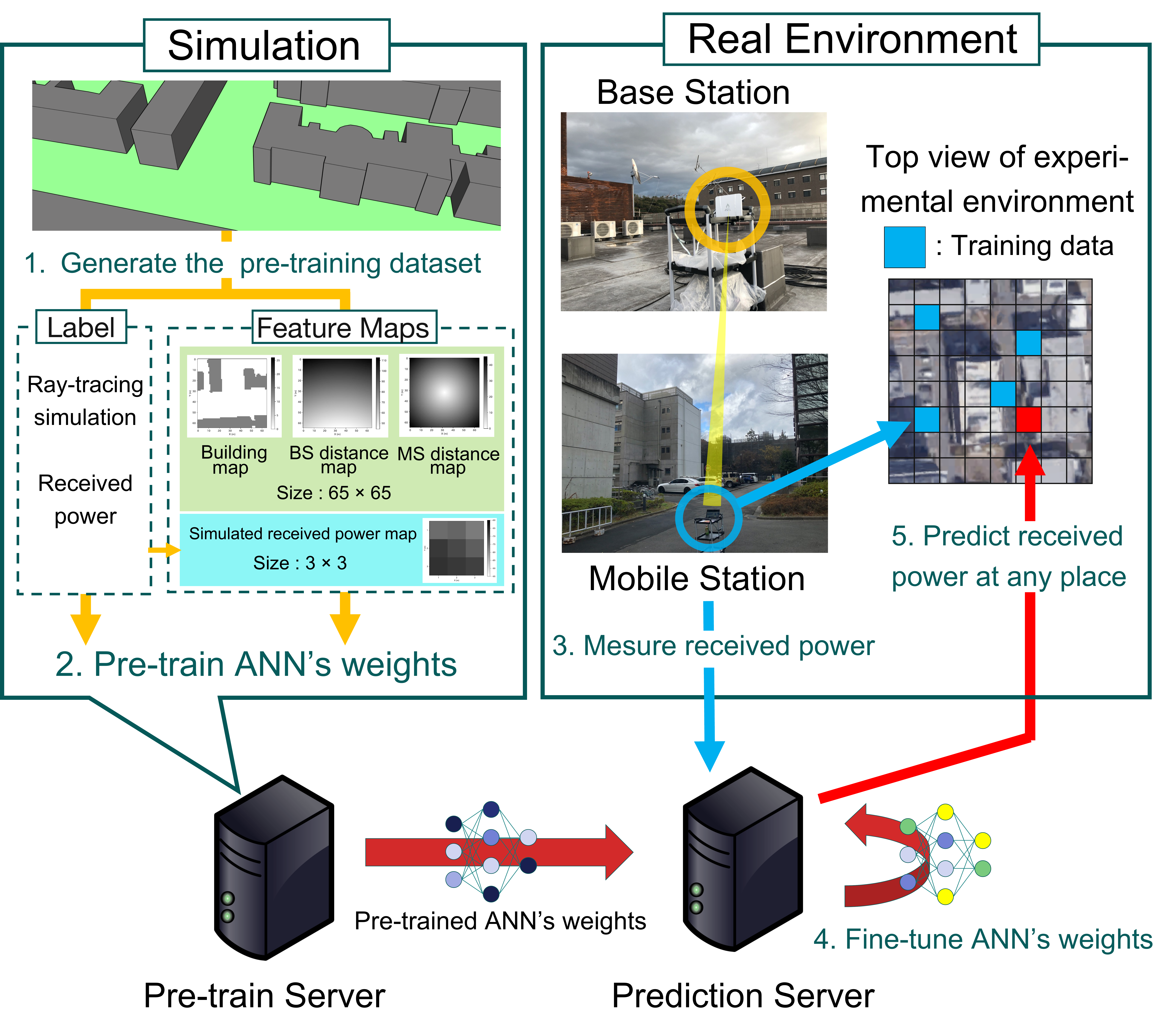}
  \caption{
    System flow. The pre-train server performs ray-tracing simulations and models the relationships between the feature maps and received power obtained from the ray-tracing simulations. The prediction server receives the received power from the MSs and updates the pre-trained weights of the ANN using small datasets measured in the real environment. The prediction server can then predict the received power at any location using feature maps.
  }
  \label{fig:system}
\end{figure}

The proposed system model is presented in \mbox{Fig. \ref{fig:system}}. This system consists of a base station (BS), mobile stations (MSs), a pre-train server, and a prediction server. The pre-train server performs ray-tracing simulations to generate a pre-training dataset. The weights of the ANN used for prediction are pre-trained using the pre-training dataset and transferred to the prediction server. The MSs send measured received power values to the prediction server with GPS position information. The prediction server then gathers feedback and generates a dataset for fine-tuning. Finally, the prediction server updates the model using the fine-tuning dataset and predicts the received power at all locations using the updated model.

\subsection{Procedures of the Proposed Method}
The proposed scheme consists of three phases: pre-training, fine-tuning, and prediction.

In the pre-training phase, a pre-training dataset is generated via ray-tracing simulations. 3D maps including buildings are used for the ray-tracing simulations. Simulations are conducted based on various scenarios (e.g., frequencies and 3D maps) that roughly corresponding to the service area for which an operator wishes to predict received power. Simulations based on a 3D map of the service area should be included because the simulation results are used to generate feature maps and perform data augmentation. The pre-train server generates a dataset from the simulation results in various scenarios including the same situation as the service area. This server then trains a model using the simulated dataset. Details regarding the input features and ANN model are presented in Sects. \ref{sec:Input} and \ref{sec:ANN}, respectively.

In the fine-tuning phase, the prediction server updates the weights of the ANN, which was pre-trained by the pre-train server, using data obtained from the service area. Small differences between simulation scenarios and the real target area are expected to be absorbed in this phase. The final training dataset is generated from received power values measured in the target service area. First, the server gathers information regarding received power values and measurement locations from the MSs. Next, the server generates an original dataset. Finally, the server augments the original dataset with the aforementioned simulation results. The details of this process are described in Sect. \ref{sec:data_aug}.

After performing fine-tuning, the prediction server predicts the received power using the fine-tuned model.

\subsection{Input Features}
\label{sec:Input}
\begin{figure}[!t]
  \centering
  \subfigure[BS distance map.]{
    \includegraphics[width=0.22\textwidth]{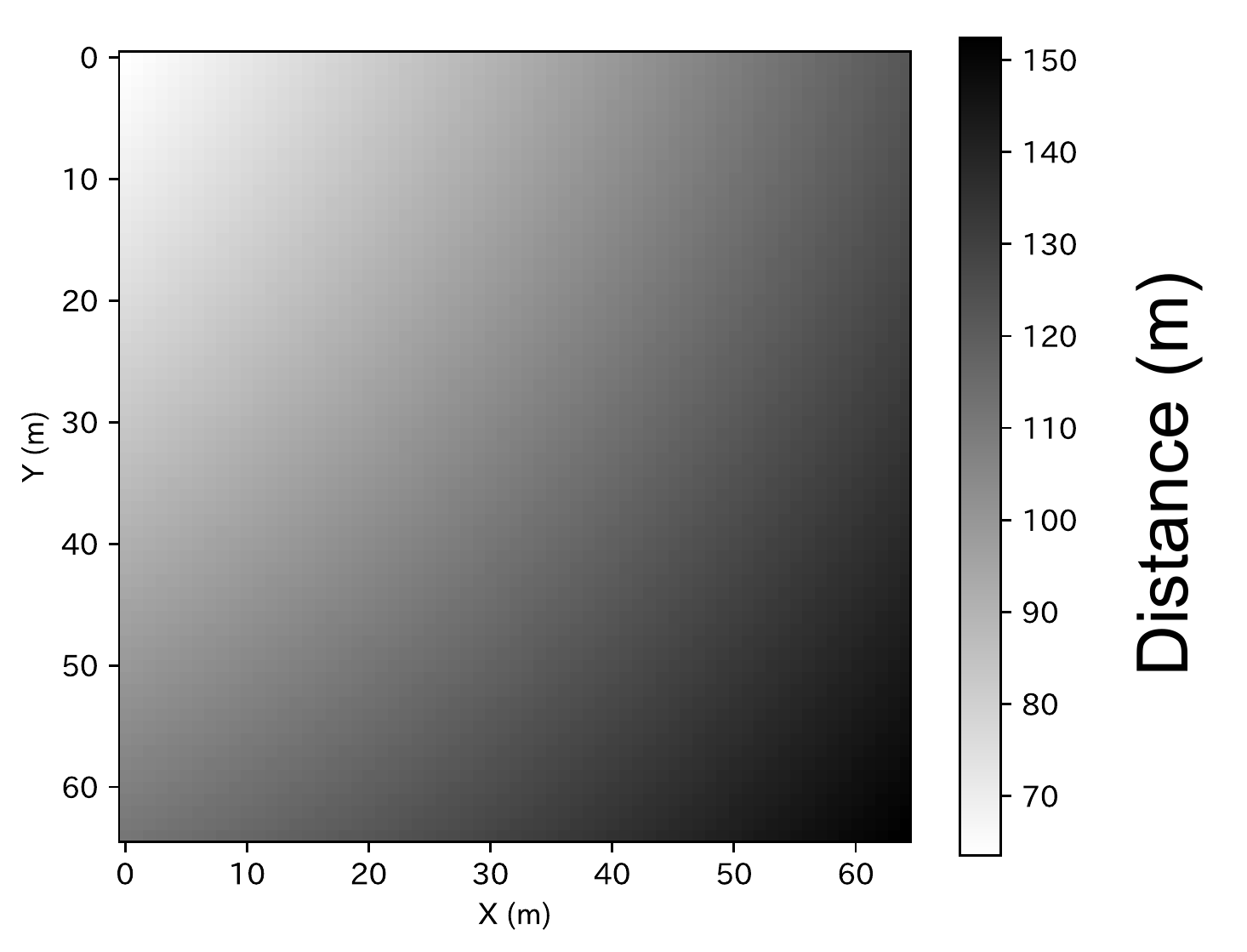}
    \label{fig:feature_BS}
  }
  \subfigure[MS distance map.]{
    \includegraphics[width=0.22\textwidth]{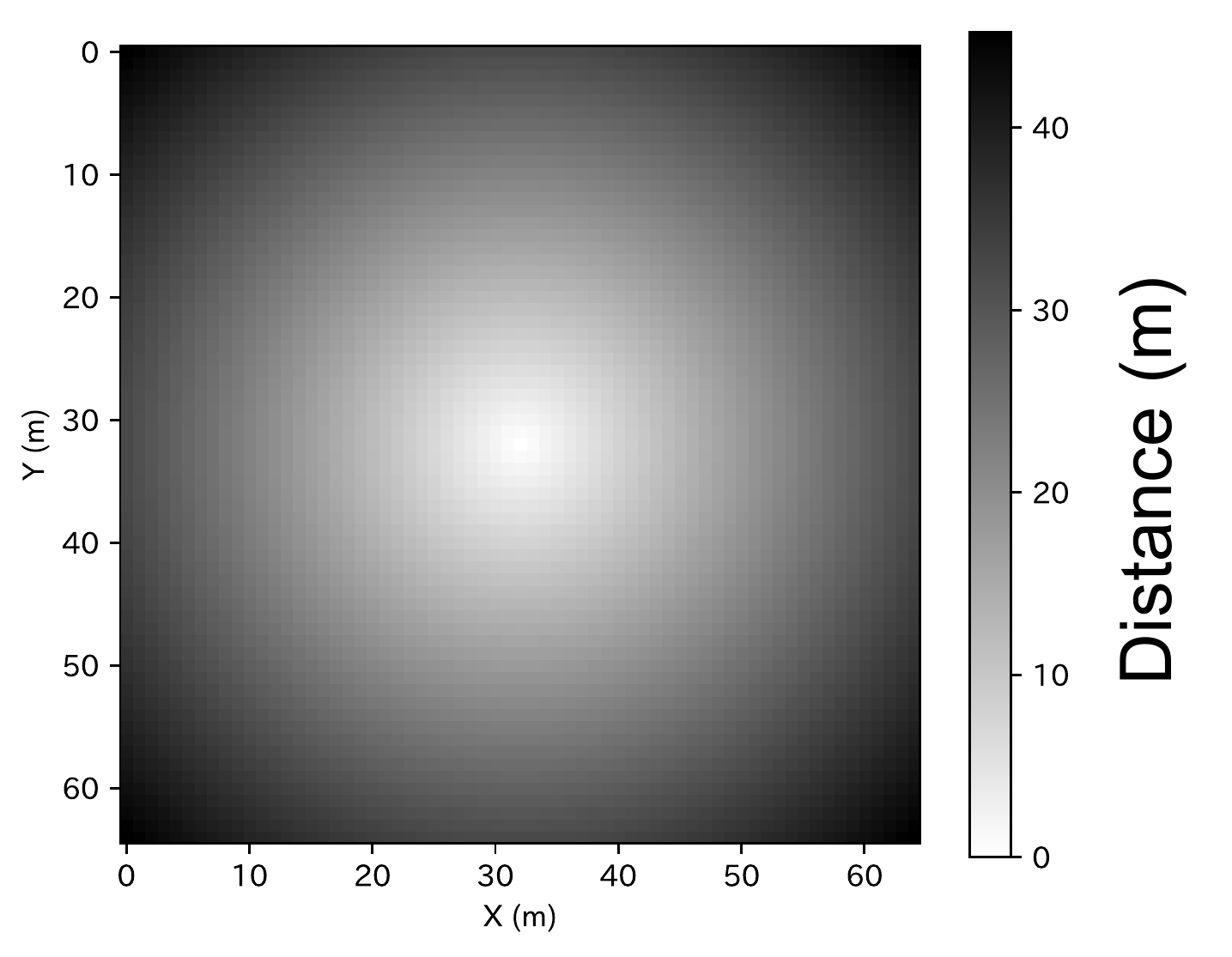}
    \label{fig:feature_MS}
  }

  \subfigure[Building map.]{
    \includegraphics[width=0.22\textwidth]{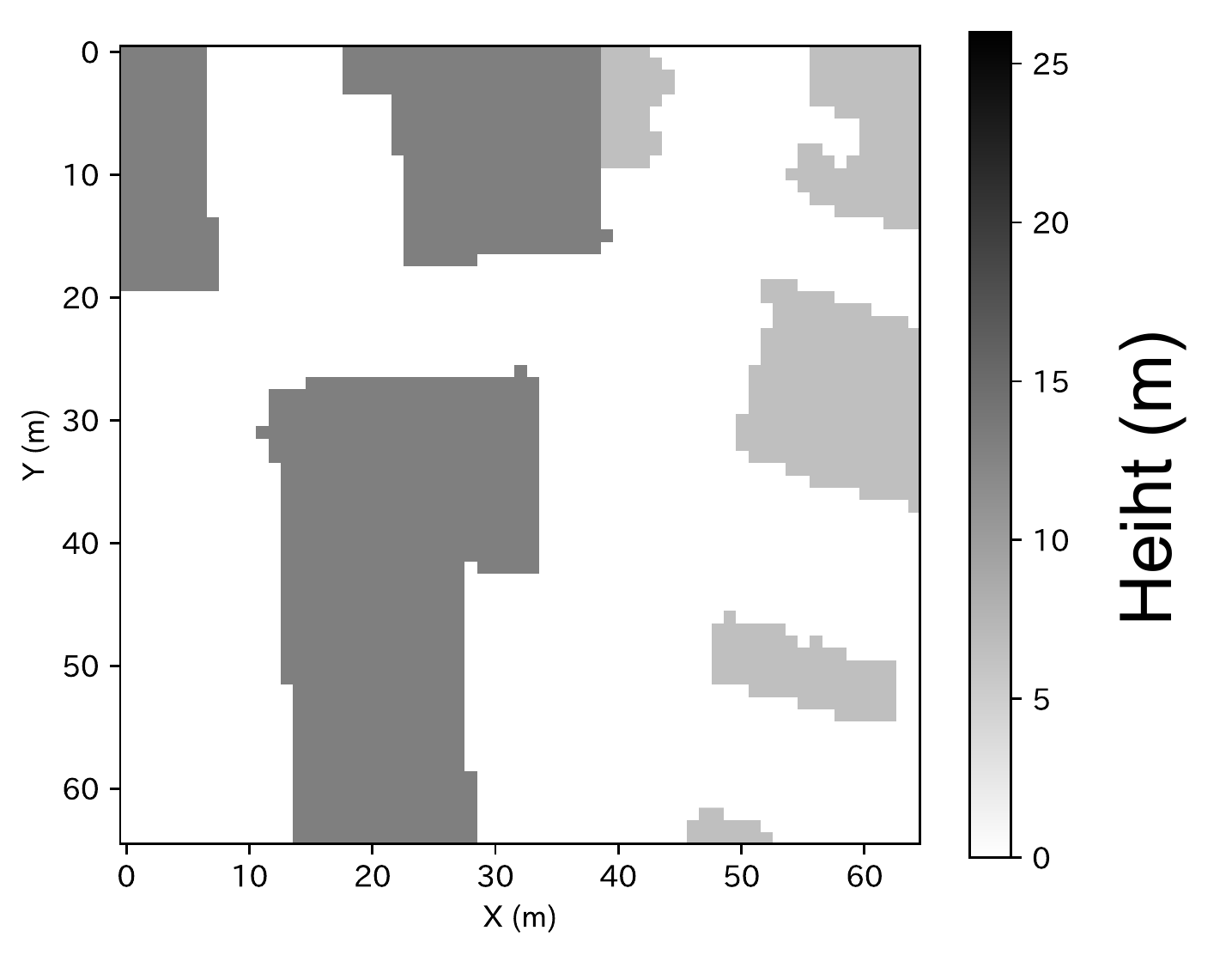}
    \label{fig:feature_building}
  }
  \subfigure[Simulated received power map.]{
    \includegraphics[width=0.22\textwidth]{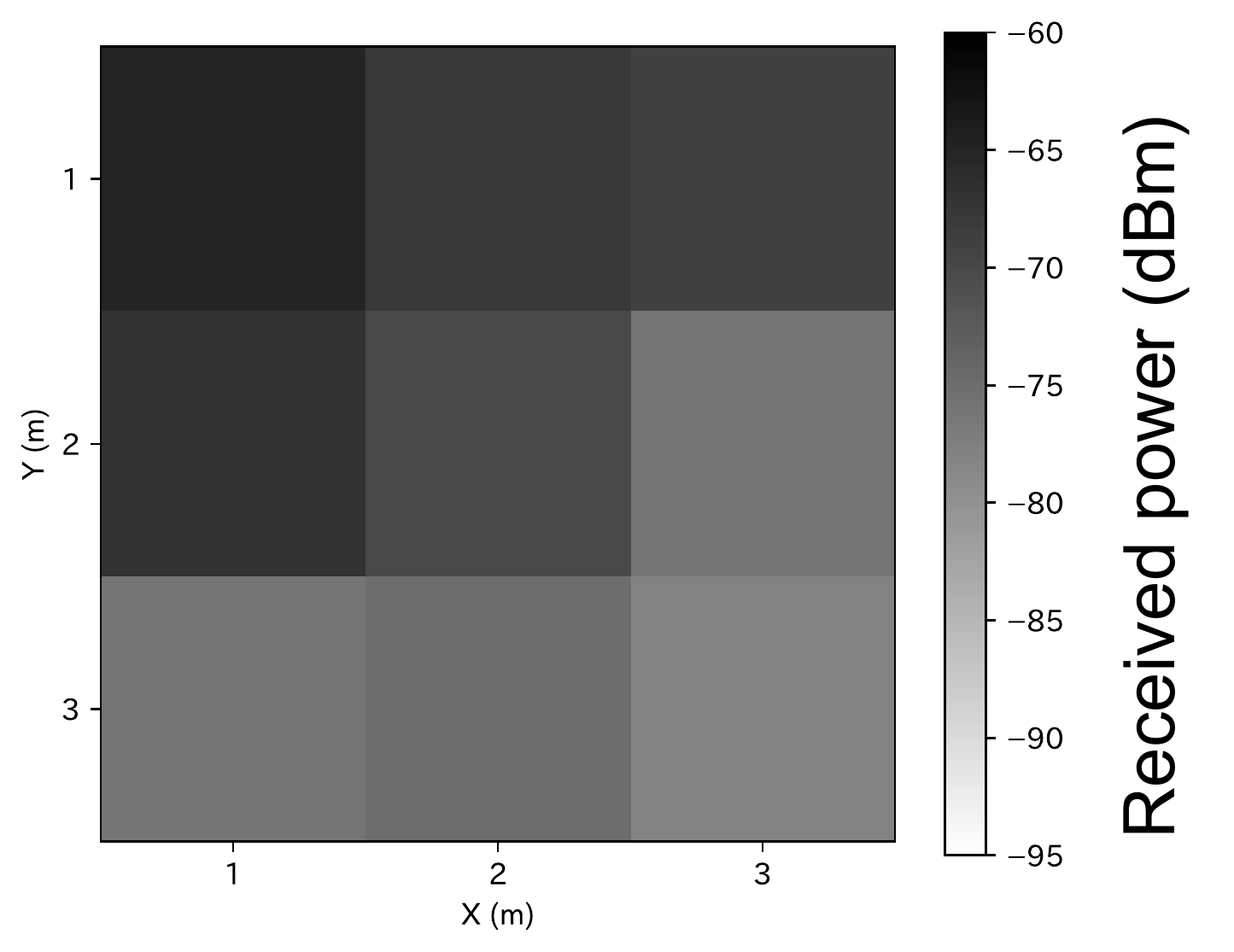}
    \label{fig:feature_RSSMAP}
  }
  \caption{Input feature maps.}
  \label{fig:feature}
\end{figure}

This section describes the features used for the received power prediction. Spatial data representing rectangular areas centered on MS locations are used as input features. \mbox{Fig. \ref{fig:feature}} presents some representative input features. In this paper, we used four image-shaped features called feature maps. Similar to a previous work \cite{imai}, \textit{BS distance map, MS distance map, and building map}, which are depicted in \mbox{Figs. \ref{fig:feature}} (a) to (c), respectively, are utilized for capturing long-term propagation characteristics (e.g., path loss and shadow fading). In the BS distance map, each grid square contains the Euclidean distance from the BS to the center of that grid square. The height of the center is set to 1\,m, which corresponds to the height from the ground to the chest of a typical mobile user. Each grid square in the MS distance map contains the distance from itself to the center of the entire grid, which represents the location of the corresponding MS. It should be noted that the MS distance map does not change when the BS and MS positions change. Each grid square in the building map contains a normalized height value relative to the ground level. This map is generated from commercially available 3D maps such as \cite{zmap}. The size of each map is 65\,$\times$\,65\,pixels, meaning each maps covers a 65\,$\times$\,65\,m area because the size of each grid square is 1\,$\times$\,1\,m.

The fourth feature is the \textit{Simulated received power map (SRP map)} shown in \mbox{Fig. \ref{fig:feature}} (d). The SRP map is included to capture the spatial correlations of received power among adjacent grid squares. Therefore, the size of this map is relatively small compared to the other feature maps, specifically, 3\,$\times$\,3\,pixels in this paper. The SRP map is generated from simulation results using a 3D map of the service area (i.e., the 3D map used to generate the building map). Specifically, the SRP map is obtained by extracting the area around each MS from the ray-tracing simulation result using the 3D map of the service area.

\subsection{ANN Configuration}
\label{sec:ANN}
\begin{figure}[!t]
  \centering
  \includegraphics[width=0.5\textwidth]{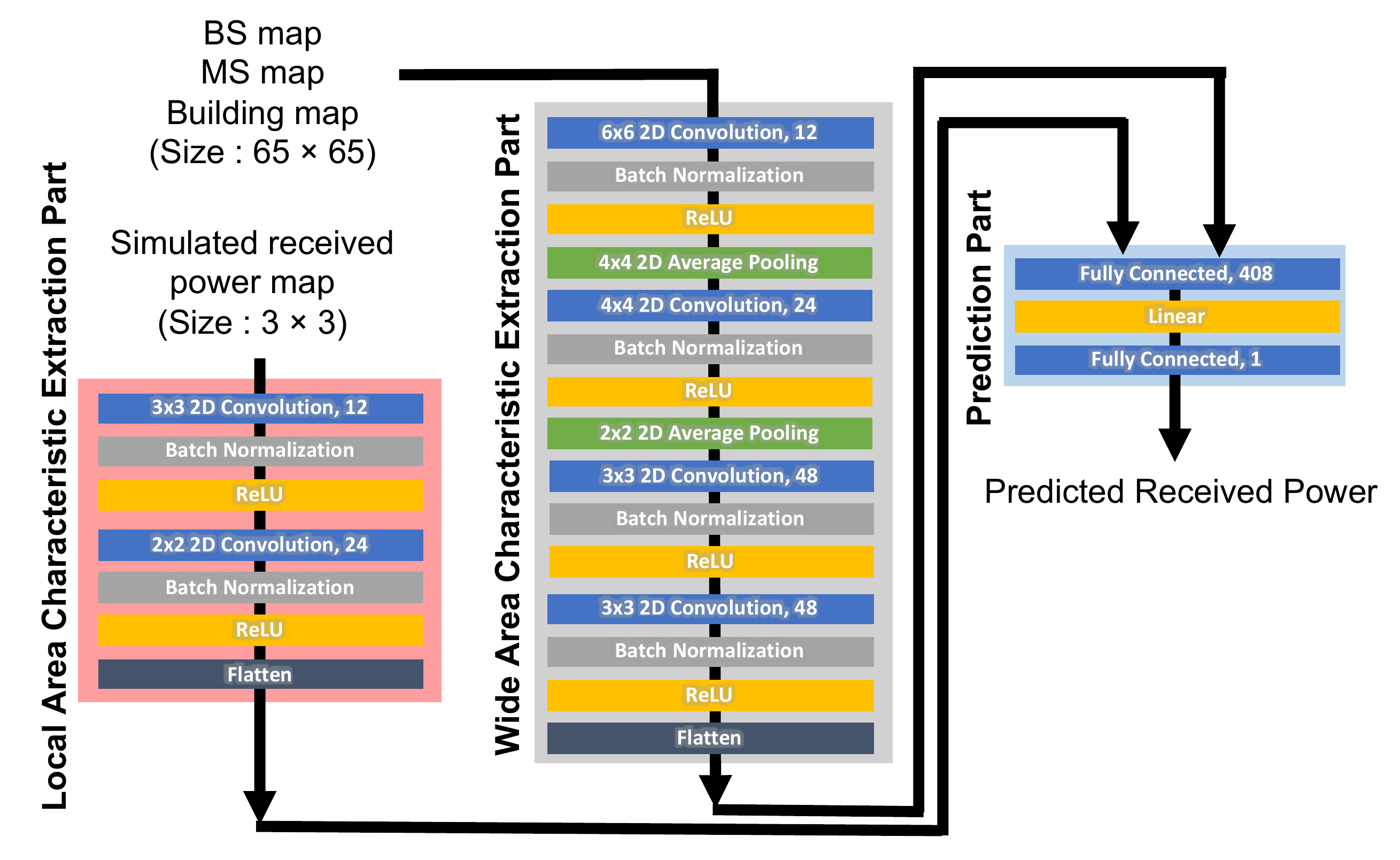}
  \caption{ANN configuration of the proposed model. The ANN is composed of a wide area characteristic extraction part, local area characteristic extraction part, and prediction part. The local area characteristic extraction part was added to the ANN presented in our previous work.}
  \label{fig:ML_model}
\end{figure}

\mbox{Fig. \ref{fig:ML_model}} shows the proposed ANN architecture. Two convolutional neural networks (CNNs) are merged into a fully connected neural network. The BS distance map, MS distance map, and building map are inputted into the larger CNN and the SRP map is inputted into the smaller CNN. This is because the sizes and objectives of these feature maps are different. Specifically, the SRP map, which is 3\,$\times$\,3\,pixels in size, is used for capturing local area characteristics. Therefore, the small CNN does not require pooling layers, which reduce the size of features and deep architectures. In contrast, the BS distance, MS distance, and building maps are 65\,$\times$\,65\,pixels in size and are utilized to capture wide area characteristics. Therefore, the corresponding CNN requires pooling layers to reduce image size and deeper layers than another one to consider relationships among distant pixels.

The architecture of each CNN refers the previous work presented in \cite{imai}. The small CNN has two convolutional layers followed by a batch normalization layer and rectified linear unit (ReLU). Batch normalization allows us to use much higher learning rates and be less careful regarding initialization \cite{pmlr-v37-ioffe15}. 
The ReLU is the most widely used activation function. 
The larger CNN consists of convolutional layers, a batch normalization layer, ReLU, and an average pooling layer. Average pooling is performed to reduce the spatial sizes of representations to reduce the associated numbers of parameters.

The final part of the proposed architecture consists of fully connected layers with a linear activation function. This part converts the features outputted by each CNN into received power at each grid square, which is the center of each feature map.

\subsection{Data Augmentation}
\label{sec:data_aug}
This section describes the proposed data augmentation method. Data augmentation is a technique that increases the size of a dataset by generating data artificially based on existing data. For image processing, data is generated by moving, rotating, and adding noise to existing images. However, augmenting wireless data is less straightforward.

As mentioned in Sect. \ref{sec:ANN}, the received power in the spatially close area has a correlation \cite{rssi_spatially}. Although the received power at a particular grid square in a simulation and real environment may differ, the spatial characteristics of the received power at surrounding grid squares are expected to be similar between a simulation and real environment. Therefore, the proposed data augmentation method generates data based on the differences between a received power value at a central grid square and those at surrounding grid squares.

\begin{figure}[!t]
  \centering
    \includegraphics[width=0.44\textwidth]{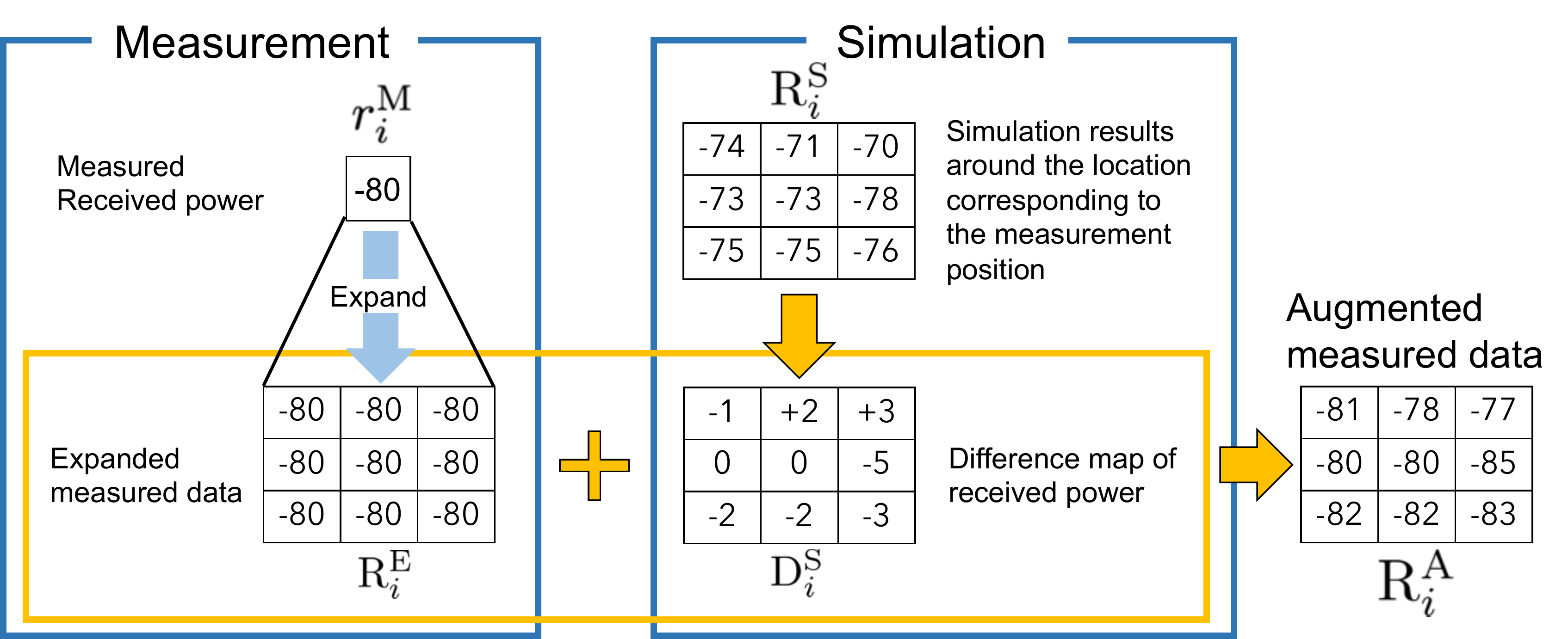}
  \caption{Proposed data augmentation method for received power prediction. When $N$ is set to 3.}
  \label{fig:data_aug}
\end{figure}

\mbox{Fig. \ref{fig:data_aug}} shows the flow of the proposed data augmentation method.
A measured received power is expanded into an $N\,\times\,N$ map.
In \mbox{Fig. \ref{fig:data_aug}}, $N$ is set to 3.
Let $r_{i}^{\rm{M}}$ denote the measured data, where $i$ identifies the grid square from which the data were obtained.
Let $\mathbf{R}_{i}^{\rm{E}}$ and $\mathbf{R}_{i}^{\rm{S}}$ denote the expanded measured data from $r_{i}^{\rm{M}}$ and ray-tracing simulation result around the measured grid.
\setlength{\mathindent}{0pt}
\begin{align}
  \mathbf{R}_{i}^{\rm{E}} &= 
  \left[
    \begin{array}{ccc}
      r_{i}^{\rm{M}} & r_{i}^{\rm{M}} & r_{i}^{\rm{M}} \\
      r_{i}^{\rm{M}} & r_{i}^{\rm{M}} & r_{i}^{\rm{M}} \\
      r_{i}^{\rm{M}} & r_{i}^{\rm{M}} & r_{i}^{\rm{M}}
    \end{array}
  \right] , \,
  \mathbf{R}_{i}^{\rm{S}} = 
    \left[
      \begin{array}{ccc}
        r_{i-4}^{\rm{S}} & r_{i-3}^{\rm{S}} & r_{i-2}^{\rm{S}} \\
        r_{i-1}^{\rm{S}} & r_{i}^{\rm{S}} & r_{i+1}^{\rm{S}} \\
        r_{i+2}^{\rm{S}} & r_{i+3}^{\rm{S}} & r_{i+4}^{\rm{S}}
      \end{array}
    \right]
  \label{ex:data_aug}
\end{align}
Then, we derive a difference map of received power values for the target measured data, as
\setlength{\mathindent}{20pt}
\begin{align}
  \mathbf{D}_{i}^{\rm{S}} &= \left[
    \begin{array}{ccc}
      r_{i-4}^{\rm{S}} - r_{i}^{\rm{S}} & r_{i-3}^{\rm{S}} - r_{i}^{\rm{S}} & r_{i-2}^{\rm{S}} - r_{i}^{\rm{S}} \\
      r_{i-1}^{\rm{S}} - r_{i}^{\rm{S}} & 0 & r_{i+1}^{\rm{S}} - r_{i}^{\rm{S}} \\
      r_{i+2}^{\rm{S}} - r_{i}^{\rm{S}} & r_{i+3}^{\rm{S}} - r_{i}^{\rm{S}} & r_{i+4}^{\rm{S}} - r_{i}^{\rm{S}}
    \end{array}
  \right] \nonumber \\
  &= \left[
  \begin{array}{ccc}
    d_{i-4} & d_{i-3} & d_{i-2} \\
    d_{i-1} & 0 & d_{i+1} \\
    d_{i+2} & d_{i+3} & d_{i+4}
  \end{array}
\right]
\end{align}
where each gird indicates a gap between the simulated received power at the gird and at the same gird as the measured data.
Finally, the difference map is added to the expanded measured data, and we obtain augmented measured data $\mathbf{R}_{i}^{\rm{A}}$.
This data augmentation method increases the number of data by $N^2$ times.
  \setlength{\mathindent}{20pt}
\begin{align}
  \mathbf{R}_{i}^{\rm{A}} &= \mathbf{R}_{i}^{\rm{E}} + \mathbf{D}_{i}^{\rm{S}} \nonumber \\
  &= \left[
    \begin{array}{ccc}
      r_{i}^{\rm{M}} + d_{i-4} & r_{i}^{\rm{M}} + d_{i-3} & r_{i}^{\rm{M}} + d_{i-2} \\
      r_{i}^{\rm{M}} + d_{i-1} & r_{i}^{\rm{M}} & r_{i}^{\rm{M}} + d_{i+1} \\
      r_{i}^{\rm{M}} + d_{i+2} & r_{i}^{\rm{M}} + d_{i+3} & r_{i}^{\rm{M}} + d_{i+4}
    \end{array}
  \right]
  \label{ex:data_aug}
\end{align}

\section{Experimental Evaluation}
\label{sec:experiment}
\subsection{Experiment Setup}
\begin{figure}[!t]
  \centering
  \subfigure[BS : Ignite Net ML-60-LW-DO]{
    \includegraphics[width=0.22\textwidth]{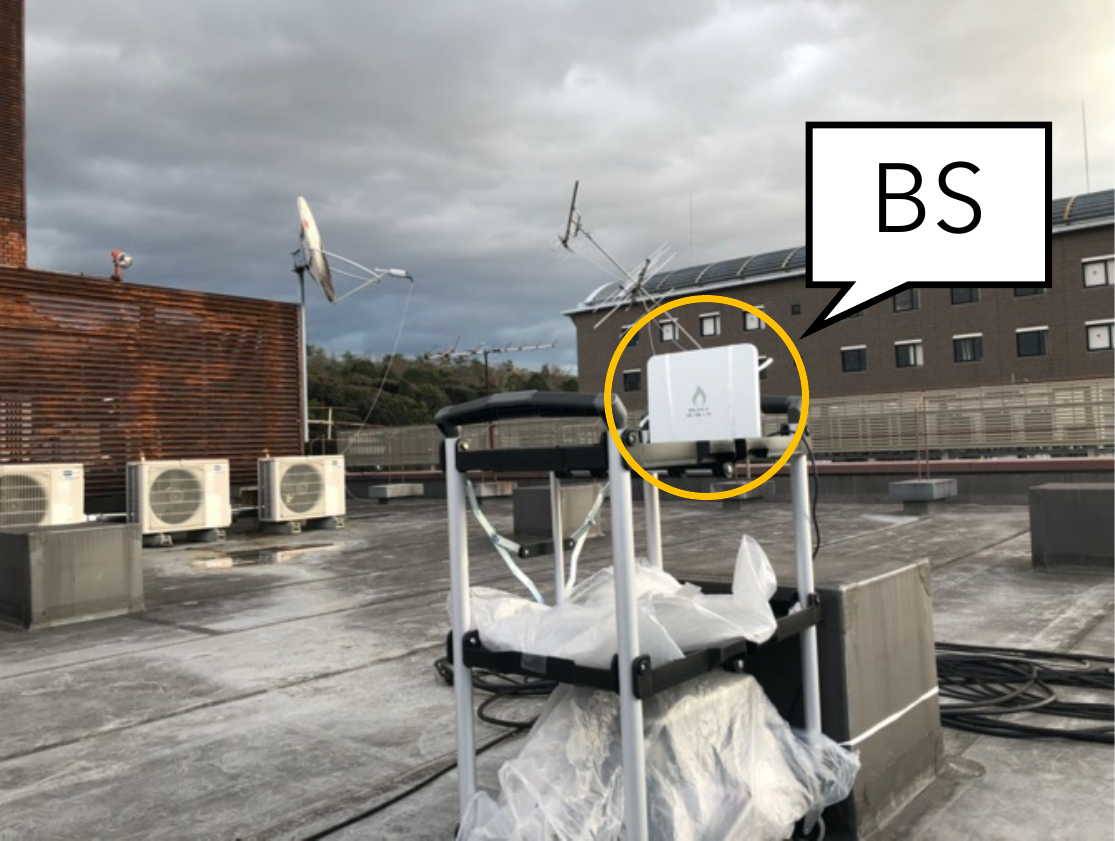}
  }\hspace{0.3em}
  \subfigure[MS : MacBook Pro 2017 13-inch]{
    \includegraphics[width=0.22\textwidth]{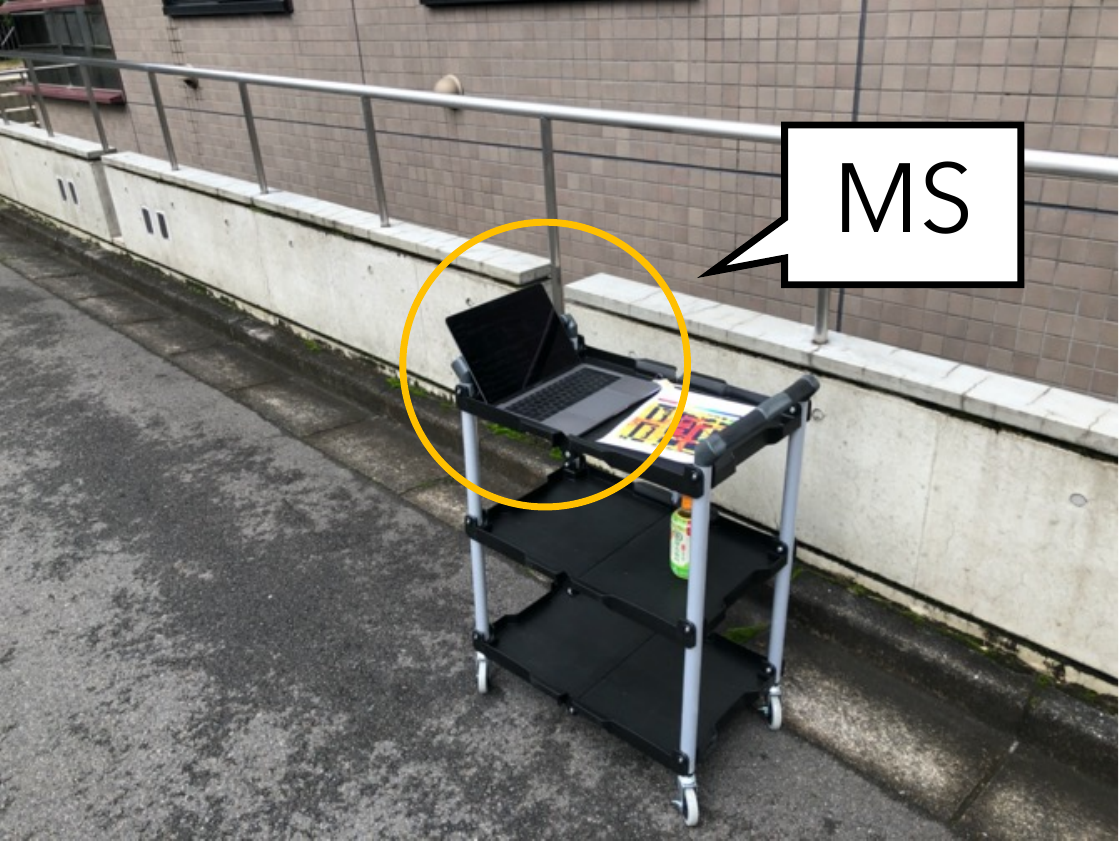}
  }
  \caption{Schematic of the experimental setup.}
  \label{fig:experimental_setup}
\end{figure}

\begin{figure}[!t]
  \centering
  \includegraphics[width=0.37\textwidth]{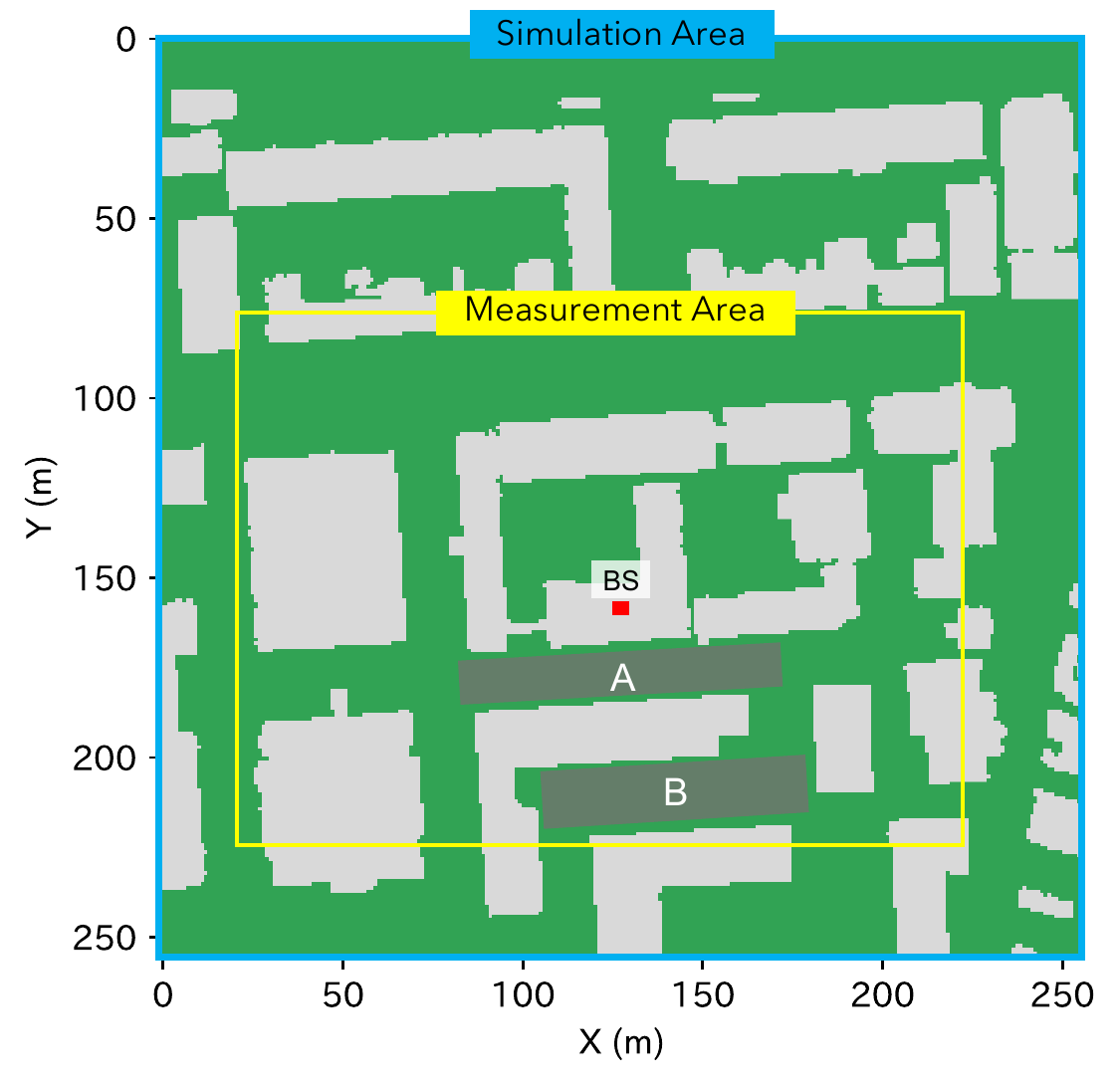}
  \caption{Simulation and measurement areas. Interpolation and extrapolation extract test data from areas ``A'' and ``B'', respectively.}
  \label{fig:Ray-tracing_area}
\end{figure}

\mbox{Fig. \ref{fig:experimental_setup}} shows the experimental setup, and the measurement specifications are listed in Table \ref{table:experimental_setup}. We conducted real-world measurements at the Yoshida Campus of Kyoto University. The measurement area is presented in \mbox{Fig. \ref{fig:Ray-tracing_area}}. The area was divided into 1\,$\times$\,1\,m grid squares. IEEE 802.11ac devices were used as both a BS and MS for measurement. The BS was installed on the roof of a building, which is indicated by the red point in \mbox{Fig. \ref{fig:Ray-tracing_area}}. The MS was located in the measurement area depicted in \mbox{Fig. \ref{fig:Ray-tracing_area}}. The BS transmitted beacon frames on channel 128, the center frequency of which is 5.64 GHz, and the MS received the frames and measured the received power. To obtain accurate received power for each position, measurements were conducted three times for each location, and the average values were recorded as the received power. We measured the average received power at 100 positions distributed uniformly within the measurement area. The values of received power ranged from -89\,dBm to -71\,dBm.

For the 100 data, we generated feature maps via ray-tracing simulations using a simulator called Wireless InSite \cite{wi} based on a 3D map called Zmap-AREAII \cite{zmap}. The simulation area is shown in \mbox{Fig. \ref{fig:Ray-tracing_area}}, where light gray blocks and green areas indicate buildings and grounds, respectively. The height values are quantized. Specifically, the height is set to zero for grounds and 6.5, 13, or 26\,m for buildings. All buildings are assumed to be made of concrete with the following dielectric half-space properties: permittivity of 6, conductivity of 0.02, and thickness of 0.3\,m. The material of the ground surface is asphalt with a permittivity of 10 and conductivity of 0.01. In our simulations, the center frequency and bandwidth were set to 5.64\,GHz and 20\,MHz, respectively. Additionally, direct, reflected, and diffracted paths were considered, but penetration paths were neglected based on the high attenuation through buildings. Details regarding the simulation settings can be found in Table \ref{table:syogen}. For the fine-tuning and performance testing, the 100 sets of measured data with feature maps were split into training data and test data with ratios of 80\% and 20\%, respectively. The training data were used to fine-tune models. The test data were used to evaluate the performance of the trained models. The test data were sampled randomly from the entire area, area A, or area B in \mbox{Fig. \ref{fig:Ray-tracing_area}}. The test data sampled from area A, which is called Data A could be more difficult than the test data sampled from the entire area since the training data does not include data obtained in area A when Data A was used. The test data sampled from area B, which is called Data B is much more difficult than Data A since the model should extrapolate for Data B.

\begin{figure}[!t]
  \centering
  \includegraphics[width=0.45\textwidth]{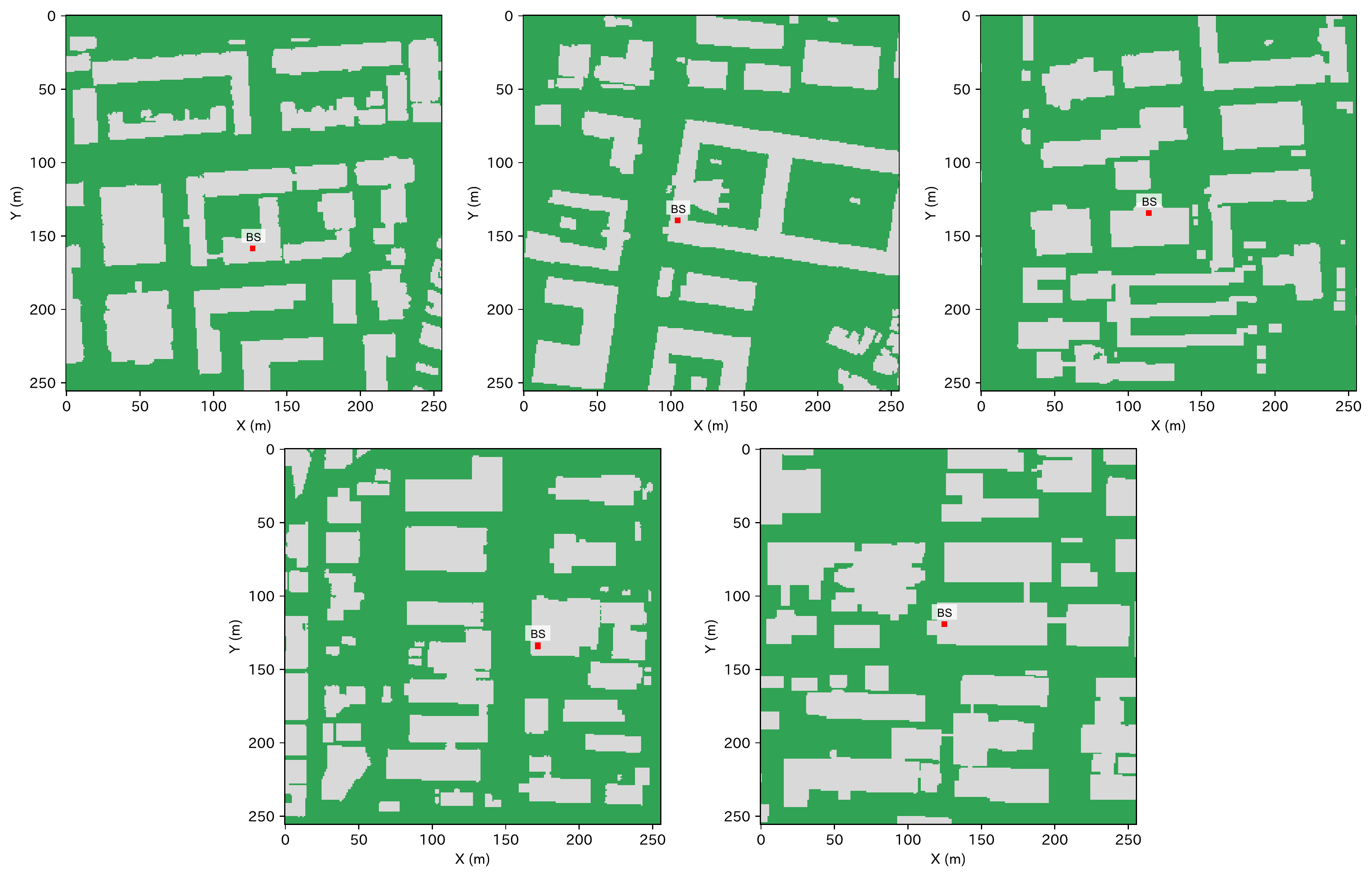}
  \caption{Simulation area in Kyoto University for generating pre-training dataset.}
  \label{fig:pretrain_area}
\end{figure}

The dataset for pre-training was generated using the ray-tracing simulation method described above in five areas in Kyoto University as shown in \mbox{Fig. \ref{fig:pretrain_area}}. One of these areas is the same area as the measurement area and the others are similar to the measurement area, which are lined with buildings about 20\,m high. We conducted the ray-tracing simulation for the simulation areas and obtained received power at each 1\,$\times$\,1\,m grid square in the areas. The total number of data samples was 107132 data samples. We randomly split the dataset into training data and test data with ratios of 80\% and 20\%, respectively, and used 25\% of the training data for validation. In the pre-training and also fine-tuning, the ANN was trained for 300 epochs using the training dataset for each.

We evaluated the root-mean-squared (RMS) error of the proposed method and the following methods for comparison: ML without pre-training, linear interpolation, kriging, and a simulation based method. Linear interpolation predicts received power by calculating linear ratios based on the surrounding four measurement points. Kriging is a geostatistical method developed for geographical interpolation and has been used successfully in a wide range of other applications, including received power prediction. The simulation based method predicts received power from simulation results by adjusting offsets. The offsets between the simulated and measured data are minimized by using the training dataset to solve the following minimization problem:
\setlength{\mathindent}{25pt}
\begin{align}
  \centering
  \min_{\alpha}\sum_{k=0}^n (y^{\rm{r}}_{k} - (y^{\rm{s}}_{k} + \alpha))^2
  \label{ex:opt}
\end{align}
where $y^{\rm{r}}_{k}$ and $y^{\rm{s}}_{k}$ represent measured values and simulation results, respectively. The position index $k$ corresponds to different positions of MSs.

In our machine learning experiments, a GeForce GTX 1080 Ti GPU was used to pre-train and fine-tune the ANNs.
An optimizer Nadam \cite{nadam} was used for the model training.

\begin{table}[!t]
  \caption{Experimental equipment.}
  \label{table:impl}
  \centering
  \renewcommand\arraystretch{1.2}
  \scalebox{0.94} {
    \begin{tabular}{ll}
      \hline
      BS & Ignite Net ML-60-LW-DO \cite{ignitenet}\\
      MS & MacBook Pro 2017 13-inch \\
      WLAN / Channel & IEEE 802.11a / 5.64\,GHz \\ 
      \hline
    \end{tabular}
  }
  \label{table:experimental_setup}
\end{table}

\begin{table}[!t]
  \caption{Simulation parameter configuration}
  \label{table:syogen}
  \centering
  \renewcommand\arraystretch{1.2}
  \scalebox{0.94} {
  \begin{tabular}{ll}\hline
    \bf{Parameter} & \bf{Value}\\ \hline
    Environment & Kyoto University\\ \hline
    Area & 256\,$\times$\,256\,\rm{m}\\ \hline
    Number of areas for pre-training & 5\\ \hline
    Max. building height & 26\,m \\ \hline
    Carrier frequency & 5.64\,GHz\\ \hline
    Bandwidth & 20\,MHz\\ \hline
    Transmit power & 21\,dBm\\ \hline
    BS altitude & 14\,m\\ \hline
    MS altitude & 1\,m\\ \hline
    Distance between adjacent MS positions & 1\,m\\ \hline
    Number of BS locations & 1\\ \hline
    Number of MS locations & 107132\\ \hline
    Max. number of reflection & 2\\ \hline
    Max. number of diffraction & 3\\ \hline
    \end{tabular}
  }
\end{table}

\subsection{Experimental Results}
\begin{figure}[!t]
  \centering
    \includegraphics[width=0.39\textwidth]{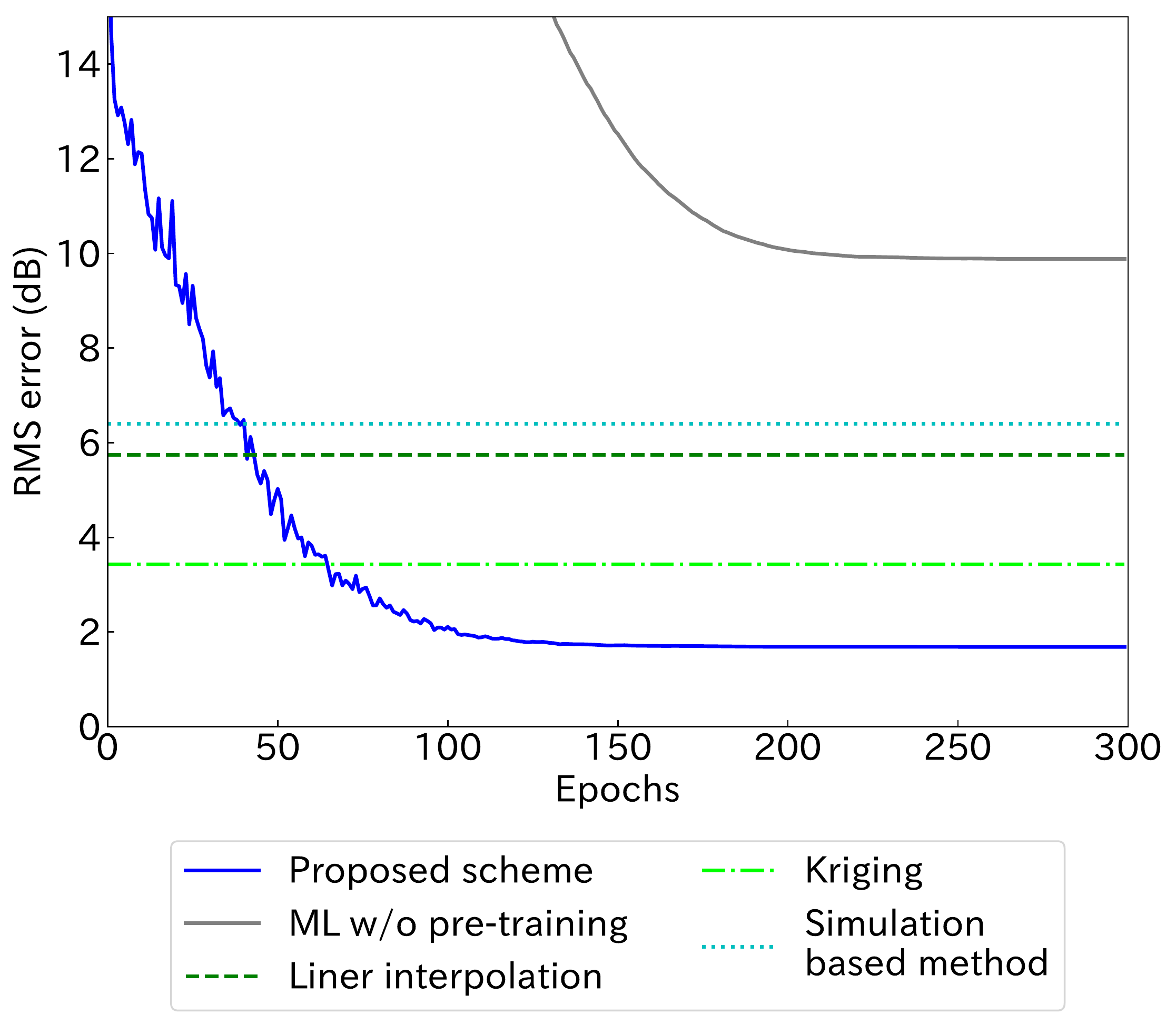}
  \caption{RMS error as a function of the number of epochs during training using the measured dataset.}
  \label{fig:result_ep}
\end{figure}

\mbox{Fig. \ref{fig:result_ep}} shows the RMS errors for randomly sampled test data. The x-axis represents the number of training epochs.
As training progresses, the RMS error of the proposed method decreases and becomes increasingly smaller than those of the other methods. 
The RMS error of the ML w/o pre-training method is much greater than those of the other methods. This is because the training data were insufficient for training an ANN from scratch. In contrast, the proposed scheme achieves low RMS error because pre-training provides much better initial values for the ANN model.

\begin{figure}[!t]
  \centering
    \includegraphics[width=0.39\textwidth]{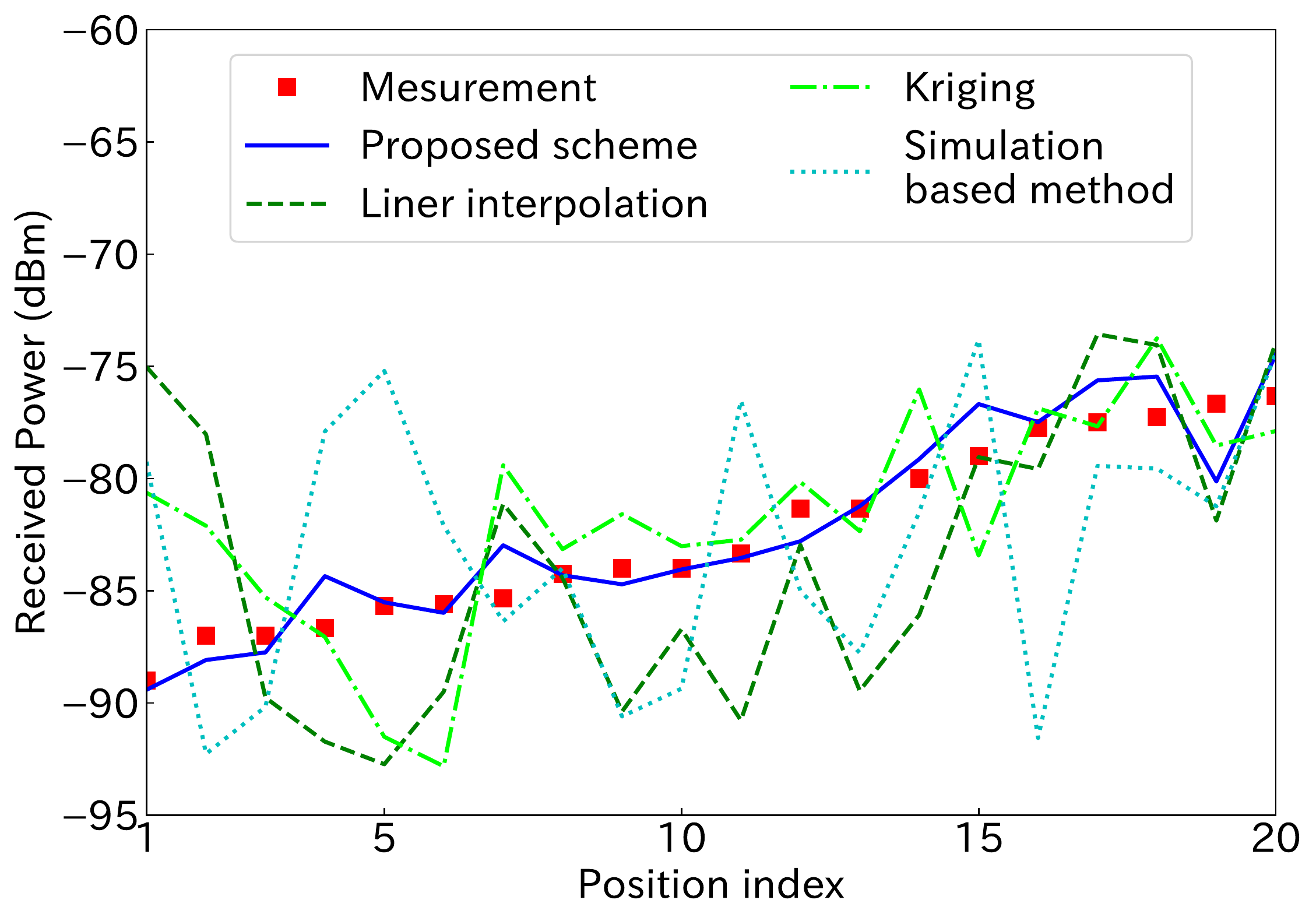}
  \caption{Measured and predicted received power values.}
  \label{fig:result_rssi}
\end{figure}

\mbox{Fig. \ref{fig:result_rssi}} shows the received power measured or predicted by the proposed and compared methods. 
The x-axis represents a position index indicating the MS positions at which the test data were measured.
The test data is sorted so that the measurement data is in ascending order.
In this evaluation, randomly sampled test data were used and data augmentation was applied for all methods.
The proposed method predicted the received power with smaller errors than the other methods, demonstrating that the proposed pre-training and fine-tuning techniques improve prediction accuracy.

\begin{figure}[!t]
  \centering
  \subfigure[Test data sampled randomly from the entire measurement area.]{
    \includegraphics[width=0.22\textwidth]{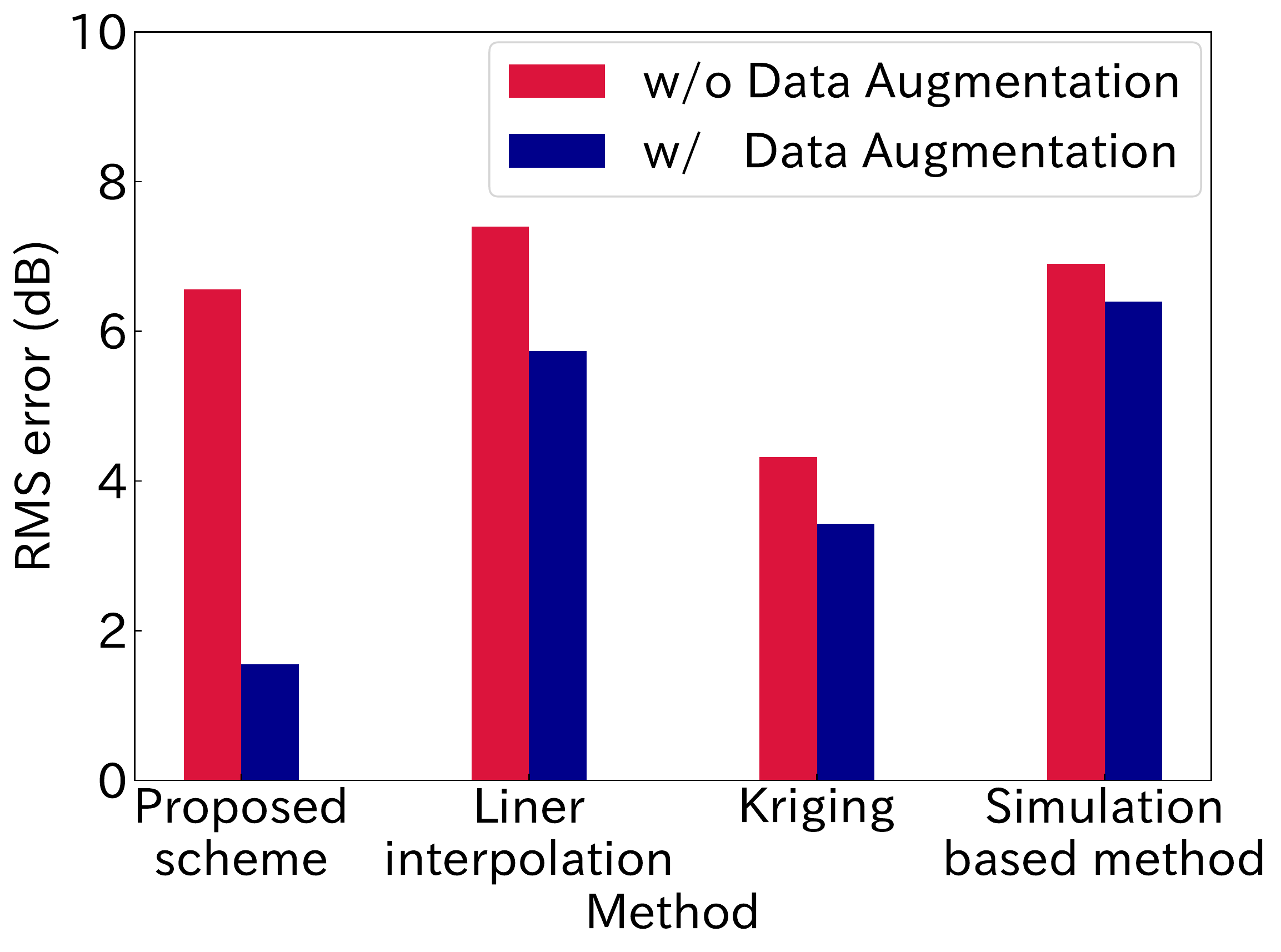}
  }
  \subfigure[Test data sampled from area A.]{
    \includegraphics[width=0.22\textwidth]{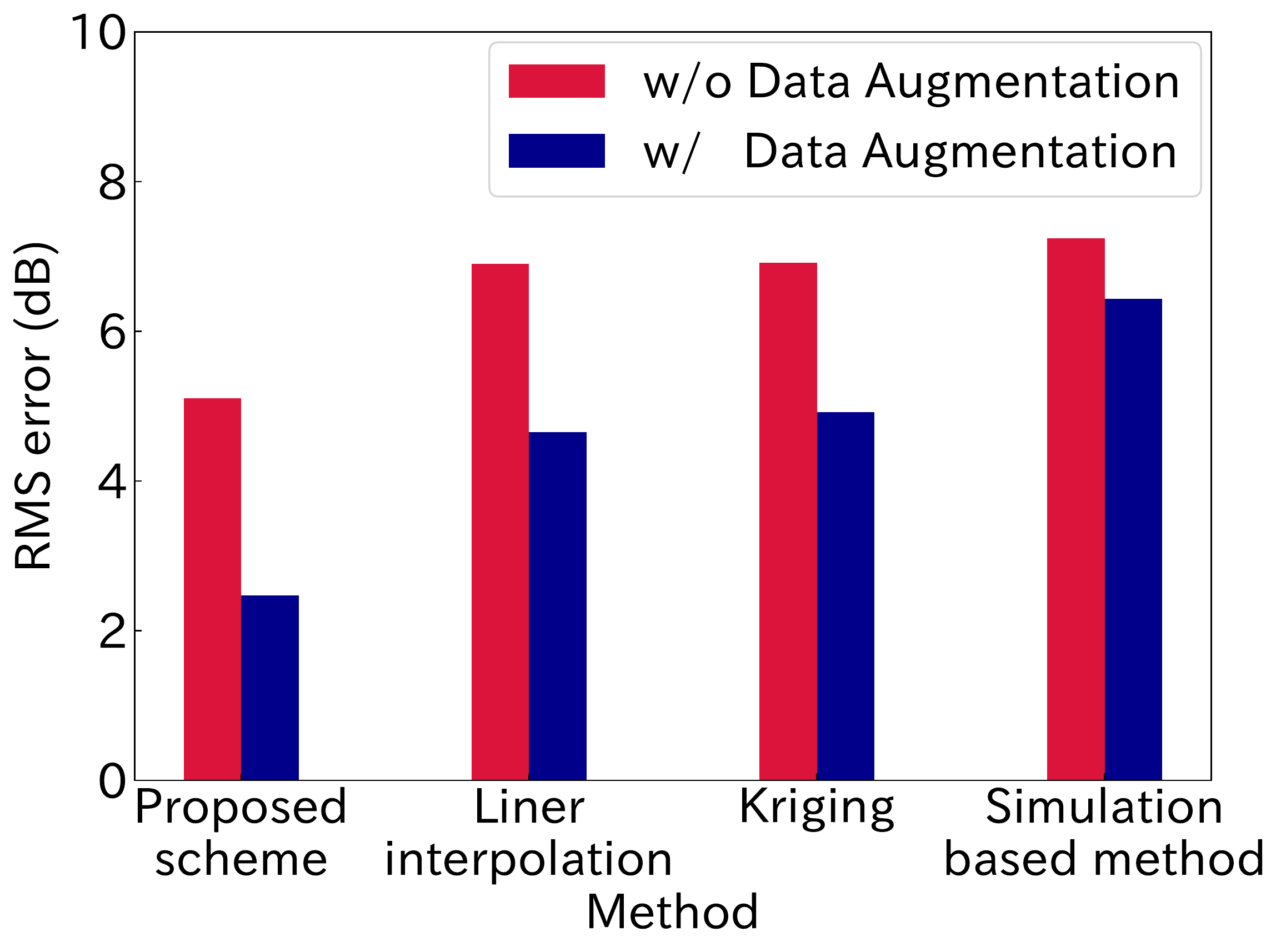}
  }

  \subfigure[Test data sampled from area B.]{
    \includegraphics[width=0.22\textwidth]{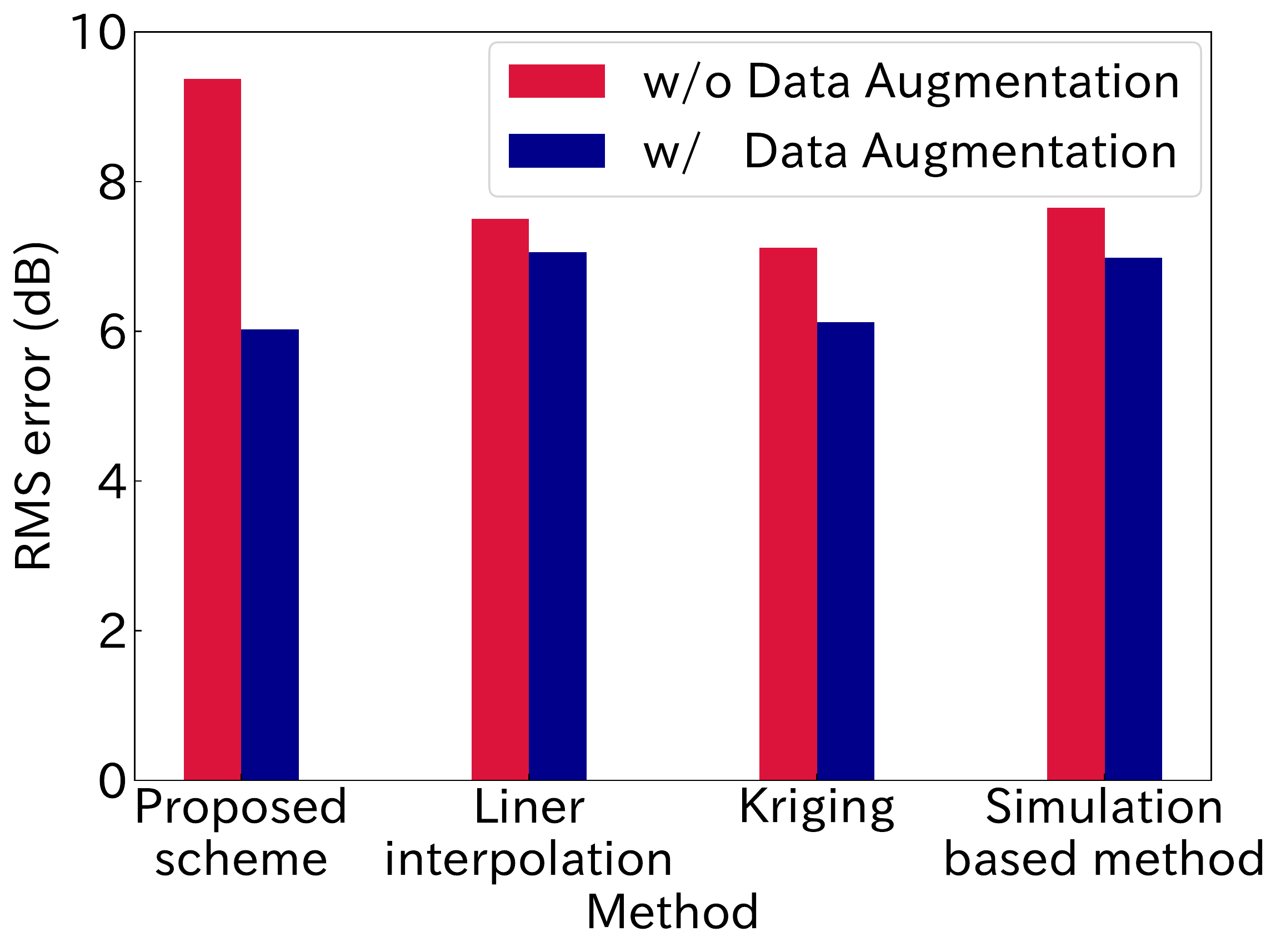}
  }

  \caption{RMS errors for test data sampled from different areas. }
  \label{fig:result_data_aug}
\end{figure}

\mbox{Figs. \ref{fig:result_data_aug}} (a), (b), and (c) show the RMS errors for test data sampled randomly from the entire area, area A, and area B, respectively.
The proposed model achieves the highest prediction accuracy for all test datasets.
Additionally, one can see that data augmentation reduces the RMS error for all of the methods. However, it is most effective for the proposed learning based method because the proposed method requires a sufficient number of data to train the ANN model.


\section{Conclusion}
\label{sec:conclusion}
This paper proposed a transfer learning-based received power prediction scheme and data augmentation method, which enabled us to train an accurate prediction model for received power prediction using a small amount of measurement data.
In the proposed method, a prediction model pre-trained using simulation data is transferred and fine-tuned using a dataset generated and augmented from small amount of data measured in a target service area using the proposed data augmentation scheme.
We experimentally confirmed that the proposed method reduces prediction error when only small amount of measurement data are available.
Specifically, the RMS error of the proposed method was reduced by more than 50\% compared to those of conventional methods.

\section*{Acknowledgment}
This work was supported in part by JSPS KAKENHI Grant Numbers JP17H03266, and JP18K13757.
  \bibliographystyle{IEEEtran}
  \bibliography{vtcf_2020}

\end{document}